\documentclass{article}
\usepackage{spconf,amsmath,graphicx}
\usepackage{url}
\urldef{\ultraxurl}\url{http://www.ultrax-speech.org}
\usepackage{booktabs}
\usepackage{gensymb}
\usepackage{csquotes}
\usepackage{hyperref}

\title{Speaker-Independent Classification of Phonetic Segments\\from Raw Ultrasound in Child Speech}
\name{Manuel Sam Ribeiro, Aciel Eshky, Korin Richmond, Steve Renals
\thanks{Supported by the EPSRC Healthcare Partnerships grant number EP/P02338X/1 (Ultrax2020 -- \ultraxurl).}}
\address{The Centre for Speech Technology Research, University of Edinburgh, UK}
\begin{document}
\ninept
\maketitle
\begin{abstract}
Ultrasound tongue imaging (UTI) provides a convenient way to visualize the vocal tract during speech production.  
UTI is increasingly being used for speech therapy, making it important to develop automatic methods to assist various time-consuming manual tasks currently performed by speech therapists.
A key challenge is to generalize the automatic processing of ultrasound tongue images to previously unseen speakers.
In this work, we investigate the classification of phonetic segments (tongue shapes) from raw ultrasound recordings under several training scenarios: speaker-dependent, multi-speaker, speaker-independent, and speaker-adapted.
We observe that models underperform when applied to data from speakers not seen at training time.
However, when provided with minimal additional speaker information, such as the mean ultrasound frame, 
the models generalize better to unseen speakers. 
%the generalization of speaker independent systems becomes easier.
%We observe that a convolutional neural network augment with speaker means achieves the best results in a speaker independent classification task.
\end{abstract}

\section{Introduction}
\label{sec:intro}
% Aciel: I think it's better to start by directly saying why we're interesting in classifying tongue shapes from ultrasound (but maybe this depends on the writing style?).
Ultrasound tongue imaging (UTI) uses standard medical ultrasound to visualize the tongue surface during speech production.
It provides a non-invasive, clinically safe, and increasingly inexpensive method to visualize the vocal tract.
Articulatory visual biofeedback of the speech production process, using UTI, can be valuable for speech therapy \cite{cleland2015using, cleland2017ultraphonix, cleland2018enabling} or language learning \cite{wilson2006ultrasound, gick2008ultrasound}.
Ultrasound visual biofeedback combines auditory information with visual information of the tongue position, allowing users, for example, to correct inaccurate articulations in real-time during therapy or learning.
In the context of speech therapy, automatic processing of ultrasound images was used for tongue contour extraction \cite{fabre2015tongue} and the animation of a tongue model \cite{fabre2017automatic}.% both of which provide visual aid for patients and therapist. 

More broadly, speech recognition and  synthesis from articulatory signals \cite{schultz2017biosignal} captured using UTI can be used with silent speech interfaces in order to help restore spoken communication for users with speech or motor impairments, or to allow silent spoken communication in situations where audible speech is undesirable \cite{denby2010silent,hueber2007eigentongue,hueber2008phone,hueber2008acquisition,hueber2010development}.
Similarly, ultrasound images of the tongue have been used for direct estimation of acoustic parameters for speech synthesis \cite{denby2004speech, csapo2017dnn, grosz2018f0}.

%such as restoring lost spoken communication, contributing to the robustness of models in noisy environments, or feedback in speech therapy and language learning .
% UTI has been used together with video images of the lips for silent speech recognition \cite{hueber2007eigentongue,hueber2008phone,hueber2008acquisition,hueber2010development}.
%Recently, both tasks were jointly modelled in a multi-task learning framework \cite{tothmulti}.
%Articulatory data such as ultrasound imaging may be used in multimodal systems to complement the speech signal.
%This can lead to more robust models, compensating for adverse noisy conditions where the audible speech signal is expected to be unreliable (schultz-2017).

Speech and language therapists (SLTs) have found UTI to be very useful in speech therapy.
In 
%the Ultrax2020 project\footnote{\url{http://www.ultrax-speech.org}}
this work we explore the automatic processing of ultrasound tongue images in order to assist SLTs, who currently largely rely on manual processing when using articulatory imaging in speech therapy.
%to alleviate some of the manual processes currently undertaken by Speech and Language Therapists (SLTs). % using audio and ultrasound.
% Aciel: some transition missing here..
One task that could assist SLTs is the automatic classification of tongue shapes from raw ultrasound.
This can facilitate the diagnosis and treatment of speech sound disorders, by allowing SLTs to automatically identify incorrect articulations, or by quantifying patient progress in therapy.  
% Aciel: I think it would be a stronger statement to say that this is (1) directly useful for therapists and (2) helps us understand the variability in ultrasound, without saying that we could have used audio but didn't. 
In addition to being directly useful for speech therapy, the classification of tongue shapes enables further understanding of phonetic variability in ultrasound tongue images. Much of the previous work in this area has focused on speaker-dependent models.  In this work we investigate how automatic processing of ultrasound tongue imaging is affected by speaker variation, and how severe degradations in performance can be avoided when applying systems to data from previously unseen speakers through the use of speaker adaptation and speaker normalization approaches.

% Steve: We can discuss this in further work
%Although admittedly these tasks can use both acoustic and ultrasound signals, we limit our analysis in this work to ultrasound data.
% This is motivated by a need to understand how automatic processing of ultrasound tongue imaging is affect by speaker variation.
% We hypothesize that a system's performance will severely degrade when applied to data from unseen speakers.
% This will illustrate the variability found in ultrasound images and motivate the need to explore adaptation or speaker normalization strategies for UTI data.

Below, we present the main challenges associated with the automatic processing of ultrasound data, together with a review of speaker-independent models applied to UTI.  Following this, we present the experiments that we have performed (Section \ref{sec:experimental_setup}), and discuss the results obtained (Section \ref{sec:results_discussion}).  Finally we propose some future work and conclude the paper (Sections \ref{sec:future_work} and \ref{sec:conclusion}).
% The following section motivates these observations by laying out some of the key challenges associated with the automatic processing of ultrasound data.
% We additionally discuss related work using speaker independent models.
% In the remainder of the paper, Section \ref{sec:experimental_setup} presents the experimental setup while Section \ref{sec:results_discussion} discusses results with respect to our initial hypotheses.
% Finally, Section \ref{sec:future_work} proposes future work and Section \ref{sec:conclusion} concludes the paper.

\begin{figure*}[ht!]
\begin{minipage}[b]{\textwidth}
  \centering
  \centerline{\includegraphics[width=\textwidth]{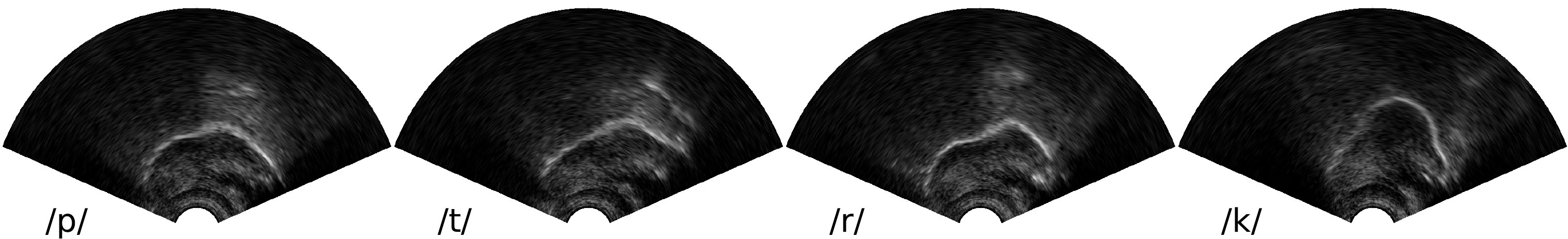}}
  %\centerline{(a) Speaker 12M}\medskip
\end{minipage}%
\vspace{.1cm}
\begin{minipage}[b]{\textwidth}
  \centering
  \centerline{\includegraphics[width=\textwidth]{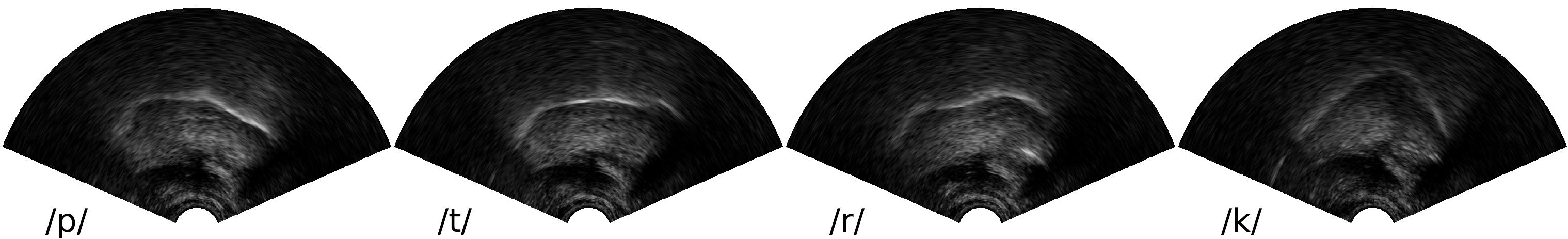}}
  %\centerline{(b) Speaker 13F}\medskip
\end{minipage} %
\caption{Ultrasound samples for the four output classes based on place of articulation. The top row contains samples from speaker 12 (male, aged six), and the bottom row from speaker 13 (female, aged eleven). All samples show a midsaggital view of the oral cavity with the tip of the tongue facing right. Each sample is the mid-point frame of a phone uttered in an aCa context (e.g. \emph{apa, ata, ara, aka}). See the UltraSuite repository\textsuperscript{2} for details on interpreting ultrasound tongue images.}
\label{fig:ultrasound}
\end{figure*}

\subsection{Ultrasound Tongue Imaging}
There are several challenges associated with the automatic processing of ultrasound tongue images.

\textbf{Image quality and limitations.} 
UTI output tends to be noisy, with unrelated high-contrast edges, speckle noise, or interruptions of the tongue surface \cite{stone2005guide, li2005automatic}. Additionally, the oral cavity is not entirely visible from the image, missing the lips, the palate, or the pharyngeal wall.

\textbf{Inter-speaker variation.} 
%is evident 
Age and physiology may affect the output, with children imaging better than adults due to more moisture in the mouth and less tissue fat \cite{stone2005guide}. However, dry mouths lead to poor imaging, which might occur in speech therapy if a child is nervous during a session. Similarly, the vocal tracts of children across different ages may be more variable than those of adults.

\textbf{Probe placement.} 
%can lead to high variability across images. 
Articulators that are orthogonal to the ultrasound beam direction image well, while those at an angle tend to image poorly. Incorrect or variable probe placement during recordings may lead to high variability between otherwise similar tongue shapes. This may be controlled using helmets \cite{spreafico2018ultrafit}, although it is unreasonable to expect the speaker to remain still throughout the recording session, especially if working with children. Therefore, probe displacement should be expected to be a factor in image quality and consistency.

\textbf{Limited data.} 
%also poses an additional challenge.
Although ultrasound imaging is becoming less expensive to acquire, there is still a lack of large publicly available databases to evaluate automatic processing methods. The UltraSuite Repository \cite{eshky2018ultrasuite}, which we use in this work, helps alleviate this issue, but it still does not compare to standard speech recognition or image classification databases, which contain hundreds of hours of speech or millions of images.

\subsection{Related Work}
Earlier work concerned with speech recognition from ultrasound data has mostly been focused on speaker-dependent systems \cite{hueber2007continuous,liu2016comparison,tatulli2017feature, ji2018updating}.
An exception is the work of Xu et al.\ \cite{xu2017convolutional}, which investigates the classification of tongue gestures from ultrasound data using convolutional neural networks.
Some results are presented for a speaker-independent system, although the investigation is limited to two speakers generalizing to a third.
Fabre et al \cite{fabre2015tongue} present a method for automatic tongue contour extraction from ultrasound data.
The system is evaluated in a speaker-independent way by training on data from eight speakers and evaluating on a single held out speaker.
In both of these  studies, a large drop in accuracy was observed when using speaker-independent systems in comparison to speaker-dependent systems.
Our investigation differs from previous work in that we focus on child speech while using a larger number of speakers (58 children).
Additionally, we use cross-validation to evaluate the performance of speaker-independent systems across all speakers, rather than using a small held out subset.
%This will illustrate the variability of results across speakers.

\section{Experimental setup}
\label{sec:experimental_setup}
\subsection{Ultrasound Data}
We use the Ultrax Typically Developing dataset (UXTD) from the publicly available UltraSuite repository\footnote{\url{http://www.ultrax-speech.org/ultrasuite}}\cite{eshky2018ultrasuite}.
This dataset contains synchronized acoustic and ultrasound data from 58 typically developing children, aged 5-12 years old (31 female, 27 male).
The data was aligned at the phone-level, according to the methods described in \cite{eshky2018ultrasuite, ribeiro2018towards}.
For this work, we discarded the acoustic data and focused only on the B-Mode ultrasound images capturing a midsaggital view of the tongue.
The data was recorded using an Ultrasonix SonixRP machine using Articulate Assistant Advanced (AAA) software at $\sim$121fps with a 135\degree \ field of view.
A single ultrasound frame consists of 412 echo returns from each of the 63 scan lines (63x412 raw frames).
For this work, we only use UXTD  type A  (semantically unrelated words, such as \emph{pack, tap, peak, tea, oak, toe}) and type B (non-words designed to elicit the articulation of target phones, such as \emph{apa, eepee, opo}) utterances.

\subsection{Data Selection}
For this investigation, we define a simplified phonetic segment classification task.
We determine four classes corresponding to distinct places of articulation.
The first consists of bilabial and labiodental phones (e.g. \emph{/p, b, v, f, \ldots/}).
The second class includes dental, alveolar, and postalveolar phones (e.g. \emph{/th, d, t, z, s, sh, \ldots/}).
The third class consists of velar phones (e.g. \emph{/k, g, \ldots/}).
Finally, the fourth class consists of alveolar approximant \emph{/r/}.
Figure \ref{fig:ultrasound} shows examples of the four classes for two speakers.

For each speaker, we divide all available utterances into disjoint train, development, and test sets.
Using the force-aligned phone boundaries, we extract the mid-phone frame for each example across the four classes, which leads to a data imbalance.
Therefore, for all utterances in the training set, we randomly sample additional examples within a window of 5 frames around the center phone, to at least 50 training examples per class per speaker. 
It is not always possible to reach the target of 50 examples, however, if no more data is available to sample from.
%Care was taken to ensure that no repeated examples were present in the training data.
This process gives a total of $\sim$10700 training examples with roughly 2000 to 3000 examples per class, with each speaker having an average of 185 examples.
Because the amount of data varies per speaker, we compute a \textquote{sampling score}, which denotes the proportion of sampled examples to the speaker's total training examples.
% ``sampling score''
We expect speakers with high sampling scores (less unique data overall) to underperform when compared with speakers with more varied training examples.

\subsection{Preprocessing and Model Architectures}
For each system, we normalize the training data to zero mean and unit variance.
Due to the high dimensionality of the data (63x412 samples per frame), we have opted to investigate two preprocessing techniques: principal components analysis (PCA, often called eigentongues in this context) and a 2-dimensional discrete cosine transform (DCT).
In this paper, \textbf{Raw} input denotes the mean-variance normalized raw ultrasound frame.
\textbf{PCA} applies principal components analysis to the normalized training data and preserves the top 1000 components.
\textbf{DCT} applies the 2D DCT to the normalized raw ultrasound frame and the upper left 40x40 submatrix (1600 coefficients) is flattened and used as input.

The first type of classifier we evaluate in this work are feedforward neural networks (\textbf{DNN}s) consisting of 3 hidden layers, each with 512 rectified linear units (ReLUs) with a softmax activation function. The networks are optimized for 40 epochs with a mini-batch of 32 samples using stochastic gradient descent. Based on preliminary experiments on the validation set, hyperparameters such learning rate, decay rate, and L2 weight vary depending on the input format (Raw, PCA, or DCT). Generally, Raw inputs work better with smaller learning rates and heavier regularization to prevent overfitting to the high-dimensional data.
As a second classifier to evaluate, we use convolutional neural networks (\textbf{CNN}s) with 2 convolutional and max pooling layers, followed by 2 fully-connected ReLU layers with 512 nodes. The convolutional layers use 16 filters, 8x8 and 4x4 kernels respectively, and rectified units. The fully-connected layers use dropout with a drop probability of 0.2.
Because CNN systems take longer to converge, they are optimized over 200 epochs.
For all systems, at the end of every epoch, the model is evaluated on the development set, and the best model across all epochs is kept.

\begin{figure}[t]
\begin{minipage}[b]{\columnwidth}
  \centering
  \centerline{\includegraphics[width=\textwidth]{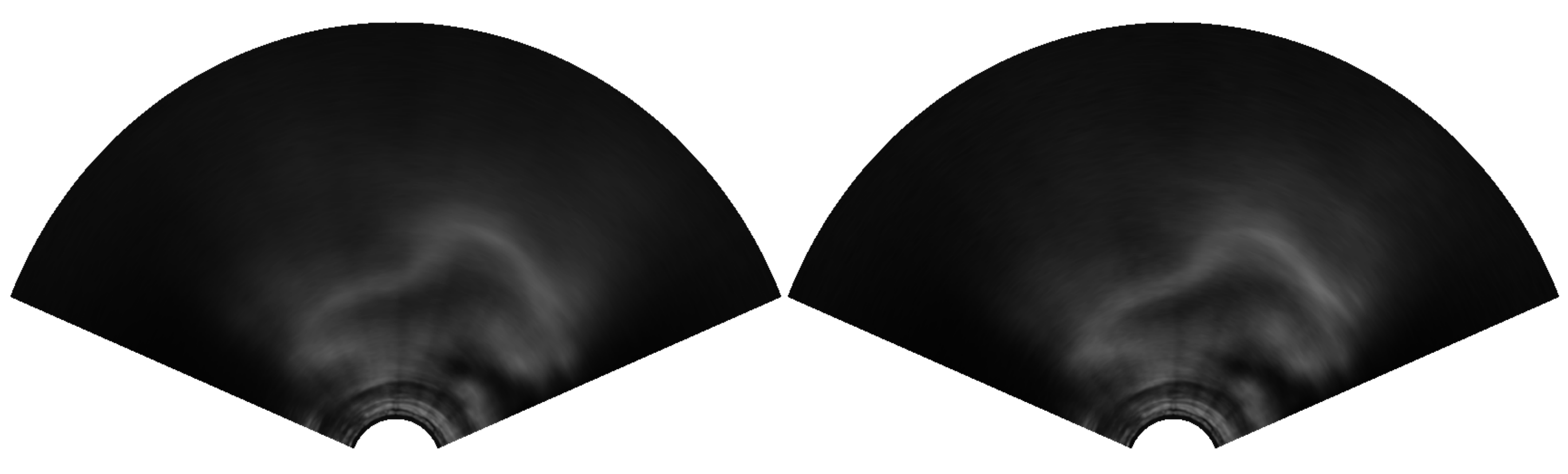}}
\end{minipage}%
\vspace{.1cm}
\begin{minipage}[b]{\columnwidth}
  \centering
  \centerline{\includegraphics[width=\textwidth]{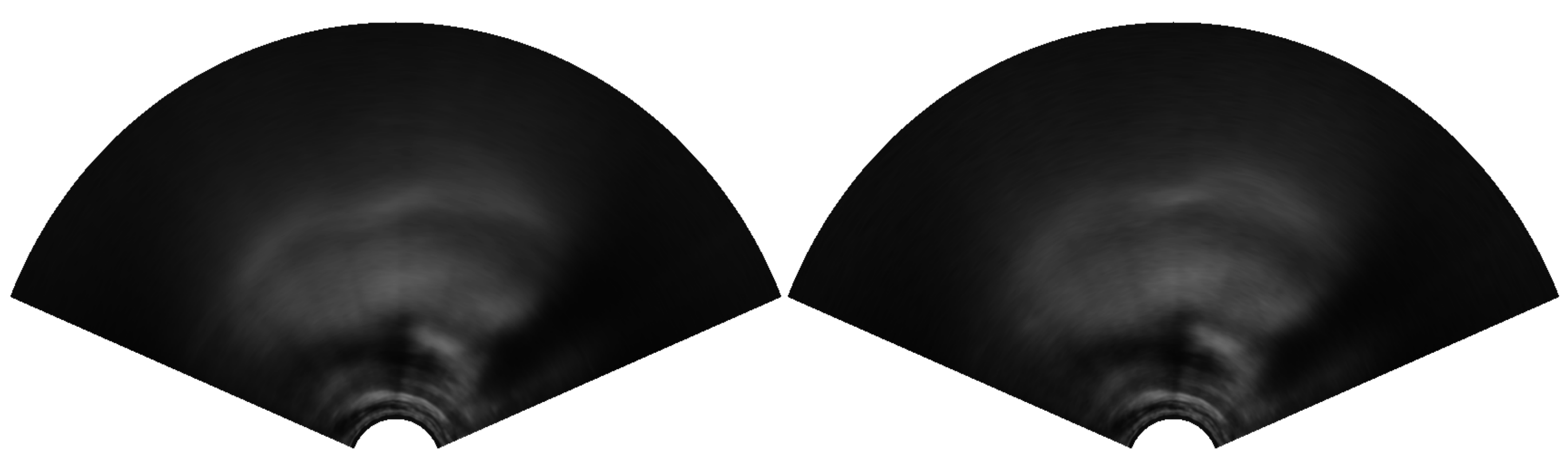}}
\end{minipage} %
\caption{Ultrasound mean image for speaker 12 (top row) and speaker 13 (bottom row). Means on the left column are taken over the training data, while means on the right are taken over the test data.}
\label{fig:ultrasound-mean}
\end{figure}

\subsection{Training Scenarios and Speaker Means}
We train speaker-\textbf{dependent} systems separately for each speaker, using all of their training data (an average of 185 examples per speaker).
These systems use less data overall than the remaining systems, although we still expect them to perform well, as the data matches in terms of speaker characteristics.
Realistically, such systems would not be viable, as it would be unreasonable to collect large amounts of data for every child who is undergoing speech therapy.
%In the figures and tables presented in this paper, we denote this system \textbf{dependent}.
We further evaluate all trained systems in a \textbf{multi-speaker} scenario.
In this configuration, the speaker sets for training, development, and testing are equal. That is, we evaluate on speakers that we have seen at training time, although on different utterances.
A more realistic configuration is a speaker-\textbf{independent} scenario, which assumes that the speaker set available for training and development is disjoint from the speaker set used at test time.
This scenario is implemented by leave-one-out cross-validation.
Finally, we investigate a speaker \textbf{adaptation} scenario, where training data for the target speaker becomes available. 
This scenario is realistic, for example, if after a session, the therapist were to annotate a small number of training examples.
In this work, we use the held-out training data to finetune a pretrained speaker-independent system for an additional 6 epochs in the DNN systems and 20 epochs for the CNN systems.
We use all available training data across all training scenarios, and we investigate the effect of the number of samples on one of the top performing systems.

This work is primarily concerned with generalizing to unseen speakers.
Therefore, we investigate a method to provide models with speaker-specific inputs.
A simple approach is to use the speaker mean, which is the pixel-wise mean of all raw frames associated with a given speaker, illustrated in Figure \ref{fig:ultrasound-mean}.
The mean frame might capture an overall area of tongue activity, average out noise, and compensate for probe placement differences across speakers.
Speaker means are computed after mean variance normalization.
For PCA-based systems, matrix decomposition is applied on the matrix of speaker means for the training data with 50 components being kept, while the 2D DCT is applied normally to each mean frame.
In the DNN systems, the speaker mean is appended to the input vector.
In the CNN system, the raw speaker mean is given to the network as a second channel.
All model configurations are similar to those described earlier, except for the DNN using Raw input. Earlier experiments have shown that a larger number of parameters are needed for good generalization with a large number of inputs, so we use layers of 1024 nodes rather than 512.

\begin{table}[t]
\centering
\resizebox{\columnwidth}{!}{%
\begin{tabular}{ccccc}
\toprule
                    & DNN Raw       & DNN PCA       & DNN DCT       & CNN Raw       \\ \midrule
Dependent           & 62.15\%       & 57.78\%       & 68.38\%       & 66.56\%       \\
Multi-speaker       & 69.62\%       & 66.30\%       & 71.91\%       & 74.70\%       \\
Independent         & 54.15\%       & 55.14\%       & 55.36\%       & 59.42\%       \\
Adapted             & 69.26\%       & 68.37\%       & 67.76\%       & 72.67\%       \\ \midrule
\multicolumn{5}{c}{with speaker mean}                                               \\ \midrule
Multi-speaker       & 71.61\%       & 67.71\%       & 72.28\%       & 74.81\%       \\
Independent         & 60.52\%       & 55.76\%       & 60.19\%       & 67.00\%       \\
Adapted             & 70.31\%       & 68.02\%       & 69.41\%       & 71.30\%       \\ \bottomrule
\end{tabular}%
}
\caption{Phonetic segment accuracy for the four training scenarios.}
\label{table:results}
\end{table}

\section{Results and Discussion}
\label{sec:results_discussion}
Results for all systems are presented in Table \ref{table:results}.
% The best performing system across training scenarios is a convolutional neural network using raw input augmented with the speaker mean.
When comparing preprocessing methods, we observe that PCA underperforms when compared with the 2 dimensional DCT or with the raw input.
DCT-based systems achieve good results when compared with similar model architectures, especially when using smaller amounts of data as in the speaker-dependent scenario.
When compared with raw input DNNs, the DCT-based systems likely benefit from the reduced dimensionality.
In this case, lower dimensional inputs allow the model to generalize better and the truncation of the DCT matrix helps remove noise from the images.
Compared with PCA-based systems, it is hypothesized the observed improvements are likely due to the DCT's ability to encode the 2-D structure of the image, which is ignored by PCA.
However, the DNN-DCT system does not outperform a CNN with raw input, ranking last across adapted systems.

\begin{figure}[t]
  \centering
  \includegraphics[width=0.9\columnwidth]{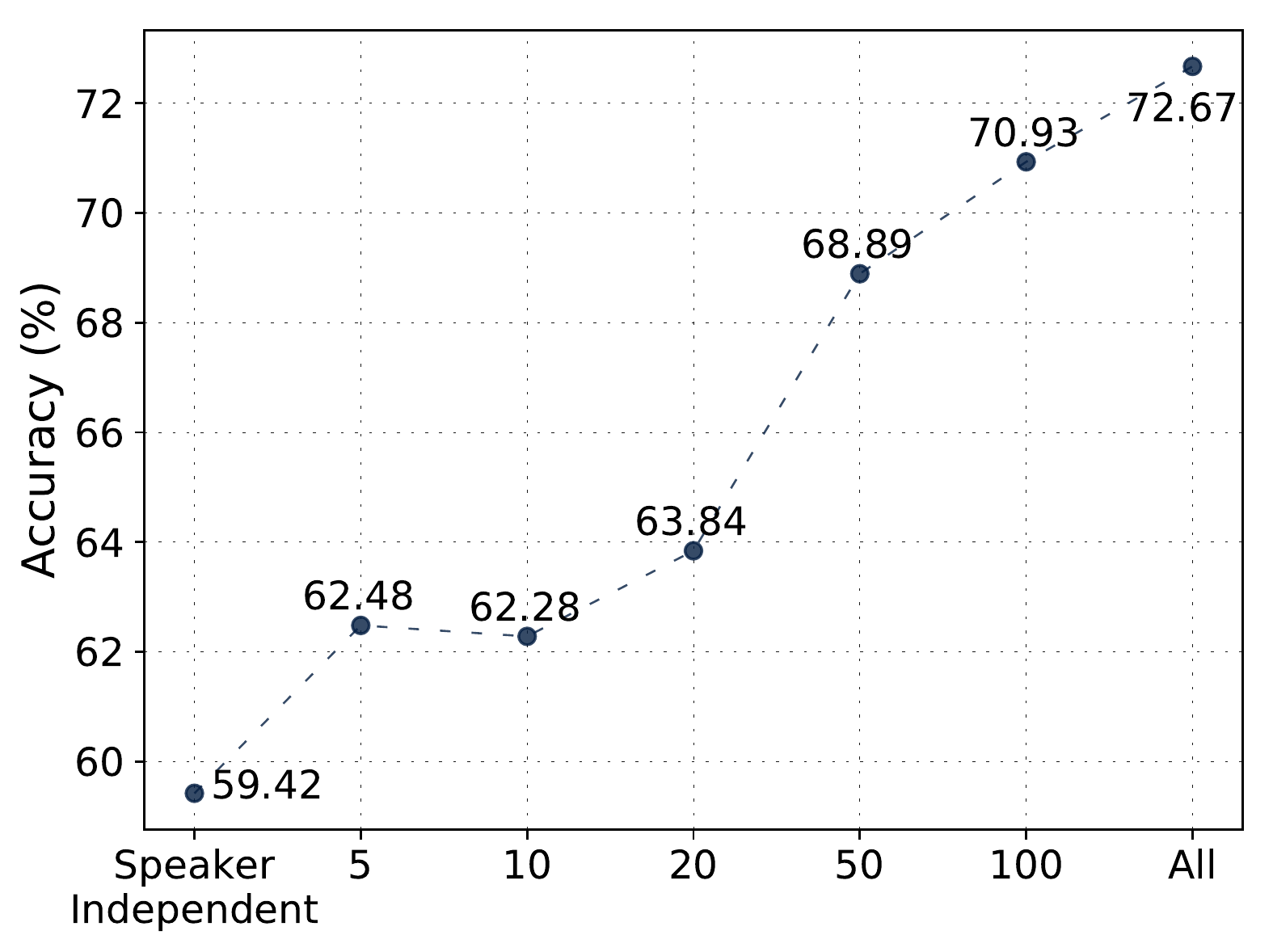}
  \caption{Accuracy scores for adapted CNN Raw, varying amount of adaptation examples. We separately restrict training and development data to either $n$ or all examples, whichever is smallest.}
  \label{fig:adapted}
\end{figure}

When comparing training scenarios, as expected, speaker-independent systems underperform, which illustrates the difficulty involved in the generalization to unseen speakers.
Multi-speaker systems outperform the corresponding speaker-dependent systems, which shows the usefulness of learning from a larger database, even if variable across speakers.
Adapted systems improve over the dependent systems, except when using DCT.
It is unclear why DCT-based systems underperform when adapting pre-trained models.
%It might be that adapting selected parts of the network could be beneficial.
Figure \ref{fig:adapted} shows the effect of the size of the adaptation data when finetuning a pre-trained speaker-independent system.
As expected, the more data is available, the better that system performs.
It is observed that, for the CNN system, with roughly 50 samples, the model outperforms a similar speaker-dependent system with roughly three times more examples.

Speaker means improve results across all scenarios.
It is particularly useful for speaker-independent systems.
The ability to generalize to unseen speakers is clear in the CNN system.
Using the mean as a second channel in the convolutional network has the advantage of relating each pixel to its corresponding speaker mean value, allowing the model to better generalize to unseen speakers.

\begin{figure}[t]
  \centering
  \includegraphics[width=0.99\columnwidth]{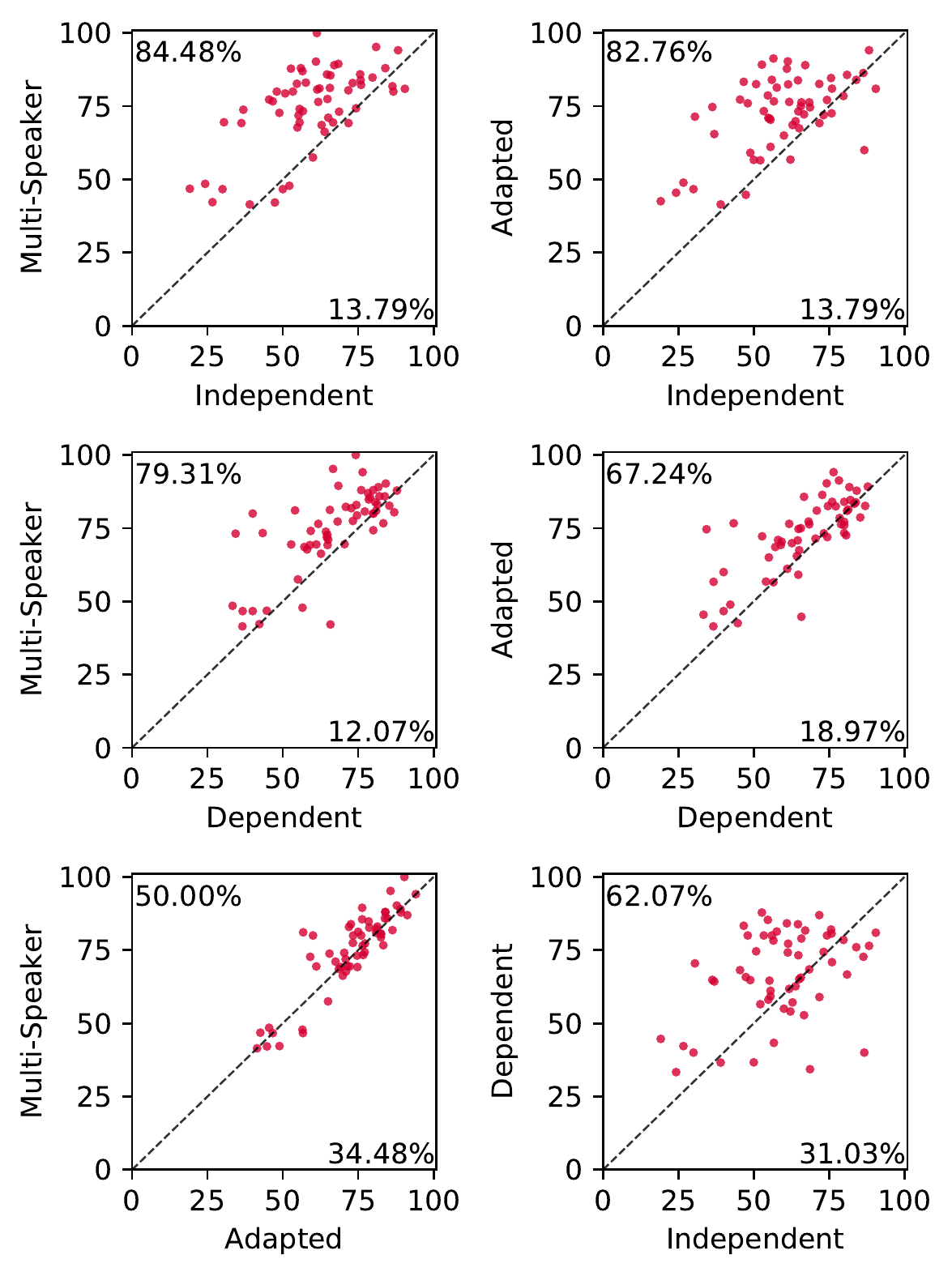}
  \caption{Pair-wise scatterplots for the CNN system without speaker mean. Each sample is a speaker with axes representing accuracy under a training scenario. Percentages in top left and bottom right corners indicate amount of speakers above or below the dashed identity line, respectively. Speaker accuracies are compared after being rounded to two decimal places.}
  \label{fig:scatter}
\end{figure}

%Overall, if using a system where no labelled data is available for new speakers, a CNN using speaker means appears to be the best solution.
%However, if additional data is available, adapting a feedforward neural network using the DCT as pre-processing step appears to be the best option.
Figure \ref{fig:scatter} shows pair-wise scatterplots for the CNN system.
Training scenarios are compared in terms of the effect on individual speakers.
It is observed, for example, that the performance of a speaker-adapted system is similar to a multi-speaker system, with most speakers clustered around the identity line (bottom left subplot).
%Similarly, a multi-speaker system is preferable to a speaker-dependent system, with most speakers above the identity line (mid-left subplot).
Figure \ref{fig:scatter} also illustrates the variability across speakers for each of the training scenarios.
The classification task is easier for some speakers than others.
In an attempt to understand this variability, we can look at correlation between accuracy scores and various speaker details.
For the CNN systems, we have found some correlation (Pearson's product-moment correlation) between accuracy and age for the dependent ($r=0.26$), multi-speaker ($r=0.40$), and adapted ($r=0.34$) systems. A very small correlation ($r=0.15$) was found for the independent system.
Similarly, some correlation was found between accuracy and sampling score ($r=-0.32$) for the dependent system, but not for the remaining scenarios. No correlation was found between accuracy and gender (point biserial correlation).

%\vspace{-0.5mm}
\section{Future Work}
\label{sec:future_work}
There are various possible extensions for this work.
For example, using all frames assigned to a phone, rather than using only the middle frame.
Recurrent architectures are natural candidates for such systems.
Additionally, if using these techniques for speech therapy, the audio signal will be available.
An extension of these analyses should not be limited to the ultrasound signal, but instead evaluate whether audio and ultrasound can be complementary.
Further work should aim to extend the four classes to more a fine-grained place of articulation, possibly based on phonological processes. Similarly, investigating which classes lead to classification errors might help explain some of the observed results.
Although we have looked at variables such as age, gender, or amount of data to explain speaker variation, there may be additional factors involved, such as the general quality of the ultrasound image.
Image quality could be affected by probe placement, dry mouths, or other factors.
Automatically identifying or measuring such cases could be beneficial for speech therapy, for example, by signalling the therapist that the data being collected is sub-optimal.

%\vspace{-0.5mm}
\section{Conclusion}
\label{sec:conclusion}
In this paper, we have investigated speaker-independent models for the classification of phonetic segments from raw ultrasound data.
We have shown that the performance of the models heavily degrades when evaluated on data from unseen speakers.
This is a result of the variability in ultrasound images,  mostly due to differences across speakers, but also due to shifts in probe placement.
Using the mean of all ultrasound frames for a new speaker improves the generalization of the models to unseen data, especially when using convolutional neural networks.
We have also shown that adapting a pre-trained speaker-independent system using as few as 50 ultrasound frames can outperform a corresponding speaker-dependent system.

%\textbf{Acknowledgements:} Supported by the EPSRC Healthcare Partnerships Programme grant number EP/P02338X/1 (Ultrax2020).

\vfill\pagebreak

% References should be produced using the bibtex program from suitable
% BiBTeX files (here: strings, refs, manuals). The IEEEbib.bst bibliography
% style file from IEEE produces unsorted bibliography list.
% -------------------------------------------------------------------------
\bibliographystyle{IEEEbib}
\bibliography{references}

%%%% Additional comments

% figure with samples from speakers with high accuracy and with low accuracy
% table mapping phones to classes
% redo experiments for speaker independent with same amount as multi-speaker
% briefly mention application of current 4-class scenario to speech therapy

\end{document}